\documentclass[twocolumn,prc,showpacs,floatfix]{revtex4}

\usepackage{amsmath,bm}
\usepackage{graphicx}

\usepackage[normalem]{ulem}  
\usepackage[dvips]{color} 

\newcommand{\vAi}{{\cal M}^{rad}_{i_1\cdots i_n}}
\newcommand{\vAim}{{\cal M}^{rad}_{i_1\cdots i_{n-1}}}
\newcommand{\vAbi}{\bar{\cal M}_{rad}^{i_1\cdots i_n}}
\newcommand{\vAbim}{\bar{\cal M}_{rad}^{i_1\cdots i_{n-1}}}

\newcommand{\htR}{\hat{R}}
\newcommand{\htB}{\hat{B}}
\newcommand{\htD}{\hat{D}}
\newcommand{\htV}{\hat{V}}

\newcommand{{\vp}}{{\vec p}}
\newcommand{{\vq}}{{\vec q}}

\newcommand{\beq}{\begin{equation}}
\newcommand{\eeq}[1]{\label{#1} \end{equation}}
\newcommand{\half}{{\textstyle \frac{1}{2}}}

\newcommand{\lton}{\mathrel{\lower.9ex
                  \hbox{$\stackrel{\displaystyle <}{\sim}$}}}
\newcommand{\ee}{\end{equation}} \newcommand{\ben}{\begin{enumerate}}
\newcommand{\een}{\end{enumerate}} \newcommand{\bit}{\begin{itemize}}
\newcommand{\eit}{\end{itemize}} \newcommand{\bc}{\begin{center}}
\newcommand{\ec}{\end{center}} \newcommand{\bea}{\begin{eqnarray}}
\newcommand{\eea}{\end{eqnarray}}
\newcommand{\beqar}{\begin{eqnarray}}
\newcommand{\eeqar}[1]{\label{#1} \end{eqnarray}}

\begin{document}
\title{ A systematic study of direct photon production in heavy ion
        collisions }

\author{Ivan Vitev$^1$}
\author{Ben-Wei Zhang$^{1,2}$}

\affiliation{$^1$ Los Alamos National Laboratory, Theoretical Division,
Mail Stop B283, Los Alamos, NM 87545, USA }

\affiliation{$^2$ Institute of Particle Physics, Hua-Zhong
Normal University, Wuhan 430079, China }

\date{\today}

\begin{abstract}

A theoretical derivation of photon bremsstrahlung, induced by the
interactions of an energetic quark in a hot and dense quark-gluon
plasma, is given in the framework of the reaction operator approach.
For the physically relevant case of hard jet production, followed by
few in-medium interactions, we find that the Landau-Pomeranchuk-Migdal 
suppression of the bremsstrahlung photon intensity is much stronger 
than in the previously discussed limit of on-shell quarks and a large 
number of soft scatterings. This result is incorporated in the first 
systematic study of direct photon production in minimum
bias d+Cu and d+Au and central Cu+Cu and Au+Au heavy ion collisions
at the Relativistic Heavy Ion Collider at center of mass energies
$\sqrt{s} = 62.4$~GeV and 200~GeV. We find that the contribution of
the photons created via final-state interactions is limited to 35\%   
for  $2$~GeV$ < p_T < 5$~GeV  and at  high transverse momenta 
the modification of the direct photon cross section is dominated by 
initial-state cold nuclear matter effects.

\end{abstract}

\maketitle

\section{Introduction}

In the highly successful hard probes program at the Relativistic
Heavy Ion Collider (RHIC), the interplay of nuclear effects
that alter the cross section for  direct  photon production is
not yet well understood. Direct $\gamma$
measurements~\cite{Adler:2005ig,Isobe:2007ku} have provided
an important baseline to help establish the dominance of
final-state effects for the observed large suppression of
energetic hadrons~\cite{Gyulassy:2003mc,:2008cx,Vitev:2005he}
in nucleus-nucleus collisions.
The interest of the theoretical community, however, has been
centered on possibly large new sources of photons as a by-product
of an energetic parton propagating in the quark-gluon plasma
(QGP)~\cite{denes}. Early work predicted significant enhancement of
low  to intermediate $p_T$ photons~\cite{Zakharov:2004bi,Turbide:2005}
based on a $\gamma$ emission pattern characteristic of on-shell
quarks and a very large number of soft interactions in matter,
e.g.~\cite{AMY:2001}.
This limit yields only a small, $\sim 30\%$, Landau-Pomeranchuk-Migdal
(LPM) suppression of the photon spectrum relative to the incoherent
Bethe-Heitler result. It does not reflect, however, the possibly
large cancellation between the bremsstrahlung associated with hard
jet production and the subsequent quark scattering, which
was found to control induced gluon emission in finite non-Abelian
plasmas~\cite{Gyulassy:2003mc,Gyulassy:2000er,Wang:2001ifa}.
Inverse Compton scattering in the QGP~\cite{Fries:2003}
has also been suggested as the dominant source of intermediate
$p_T$  photons if the $\gamma$ acquires most of the momentum of the
incoming quark. Such enhancement effects, in conjunction with their
associated  negative azimuthal  asymmetry coefficient
$v_2(p_T)$~\cite{Turbide:2005,Turbide:2005bz}, do not appear to
be compatible with existing direct photon
data at RHIC~\cite{Isobe:2007ku,Pantuev:2007,takao}.

Recent phenomenological refinements~\cite{Arleo:2007bg,Turbide:2007mi}
suggest that the QGP enhancement of direct $\gamma$ production
may be smaller than previously expected and partly cancelled by the
quenching of fragmentation photons. Still, there is no calculation to date
that consistently includes known nuclear matter effects, such as the
Cronin  effect~\cite{Accardi:2002ik,Vitev:2003xu,Qiu:2003pm},
shadowing~\cite{Eskola:1998df,Qiu:2004da}, and cold
nuclear matter energy loss~\cite{Vitev:2007ve}, to provide
quantitative guidance for the relative strength of initial- and
final-state modifications in the observed $\gamma$ cross sections.
An additional serious deficiency in the theory and phenomenology of direct
photon production in heavy ion collisions is the absence
of systematic studies in proton-nucleus (p+A) and nucleus-nucleus (A+A)
collisions for different system sizes and center of mass energies.
This is especially true now, when new experimental results from RHIC
are soon expected to become available~\cite{takao,Peressounko:2006qs}.
Last but not least, only through extensive detailed comparison
between theory and data~\cite{:2008cx} can one test the model
validity and gain confidence in the extracted quantitative
properties~\cite{Adare:2008cg} of the dense matter created in
heavy ion reactions.

With this motivation, we derive the QGP-induced $\gamma$ spectrum
for hard quark production in finite size plasmas. The same model of
jet-medium interactions is used to calculate the quark conversion cross
section and the suppression of fragmentation photons. These theoretical
results, when applicable, are combined in a numerical simulation
with cold nuclear matter effects to provide model predictions for
d+Cu, d+Au, Cu+Cu and Au+Au reactions  at center of mass energies
of 62.4~GeV and 200~GeV per nucleon pair at RHIC.
This article is organized as follows:  in section~\ref{zero} we
highlight the differences between gluon and photon bremsstrahlung
and identify the theoretical approach that reproduces the known
Bethe-Heitler spectrum. The derivation of the final-state
medium-induced photon radiation in the reaction  operator approach
is given in section~\ref{allord}. Numerical results, relevant to
the phenomenology of direct $\gamma$ production are also shown.
In section~\ref{study} we carry out a systematic investigation of
cold and hot nuclear matter effects that alter the mid-rapidity photon
cross section in ultra-relativistic collisions of heavy nuclei at
RHIC. A summary and conclusions are presented in
section~\ref{conclude}.

\section{Photon versus gluon bremsstrahlung}
\label{zero}

The computation of photon bremsstrahlung is usually considered
to be easier than that of gluon bremsstrahlung due to the absence
of self-interactions of the gauge boson. However, it is not well
appreciated that the two physics processes are quite different.
To illustrate this, we first examine the radiative amplitude
that corresponds to the case of single scattering of a fast  on-shell
quark~\cite{Gunion:1981qs,Gyulassy:2000er}:
\beqar {\cal M}_{rad}(k)
\propto  2  i g_s {\bf \epsilon}_\perp \cdot \bigg( \frac{ {\bf
k}_\perp}{{\bf k}_\perp^2}  - \frac{ ({\bf k}-{\bf q})_\perp}{({\bf
k}-{\bf q})_\perp^2 } \bigg) e^{i\frac{{\bf k}_\perp^2}{2k^+} z^+ }
[T^c,T^a] \; .  \;
\eeqar{bg}
In Eq.~(\ref{bg}) $g_s$ is the strong
coupling constant, $k^\mu = [k^+,k^-, {\bf k}_\perp ]$ is the
momentum of the radiated gluon in light-cone coordinates and
$\epsilon$ is its polarization vector. We denote by  $q^\mu = [
q^+=0,q^-,{\bf q}_\perp ]$ the momentum exchange with the medium at
position $z$ and by $T^c,T^a \in {\rm SU(3)}$  the color matrices
at the emission and interaction vertexes. Evidently, it is the color
rotation of the parent parton and the re-interaction of the bremsstrahlung
gluon in nuclear matter that determine the gluon emission intensity
and allow neglect of the deflection of the jet.
If we, however, take the Quantum Electro-Dynamics (QED)
limit $g_s \rightarrow e_q =(\pm 1/3, \pm
2/3)e$, $ T_c \rightarrow {\bf 1} $ in Eq.~(\ref{bg}) we find
${\cal M}_{rad}^\gamma(k) \rightarrow 0 $. Therefore, a theoretical
approach developed to describe gluon emission cannot be directly
generalized to photon emission and vice versa~\cite{Zhang:2003}.
Our conclusion is independent of the specific example of incoherent
parton scattering. All  regimes of coherent inelastic scattering
can be treated in the unified framework of the reaction operator
approach~\cite{Vitev:2007ve}. The reaction operator $\htR_n$
describes the effect of one additional correlated in-medium scattering
at position $z_n$ at the cross section level and is process dependent.
Taking the above mentioned QED limit of $\htR_n$ derived for gluon
bremsstrahlung~\cite{Gyulassy:2000er}, one finds:
\beqar \htR_n =
T_a T_a - (C_F/2) {\bf 1} - (C_F/2) {\bf 1} \equiv {\bf 0} \;.
\eeqar{zrop}
In Eq.~(\ref{zrop}) $C_F$ is the quadratic Casimir in
the fundamental representation of SU(3). Thus, better treatment of
jet-medium interactions is needed for both incoherent and coherent
photon emission calculations.

With these results in mind, we first identify the refinement
of the kinematic approximations necessary to derive the induced
$\gamma$ spectrum. The scattering of a fast quark in nuclear matter
is  modelled via interactions with  an external non-Abelian field
$V^{\mu,c} (q)$~\cite{Qiu:2003pm}: \beqar V^{\mu,c} (q) &=&
n^{\mu}\, 2\pi
\delta(q^+)\, V^c (q)\, e^{i q \cdot z} \;, \nonumber \\
 g_s V^c(q) &\equiv&  v(q) \, T^c(t) \;\;.
\eeqar{V-solution}
Here, the four-vector $n^\mu = \delta^{\mu,-} = [0,1,{\bf 0}_\perp ]$ and
the color matrix $T^c(t)\in {\rm SU_c(3)}$ in Eq.(\ref{V-solution})
represents the target charge that creates the non-Abelian
field. We take the Fourier transform $v(q)$ to be of color-screened
Yukawa type but with Lorentz boost invariance:
\beq
v(q) \equiv  \frac{4\pi \alpha_s}{-q^{\;2}+\mu^2}
= \frac{4\pi \alpha_s}{{\bf q}_\perp^{\;2}+\mu^2}
 = v({\bf q}_\perp) \;\; ,
\eeq{vq}
where we have used the $q^+=0$ choice of frame. This specific form
of $v(q)=v({\bf q}_\perp)$ is particularly useful since in-medium
interactions in both hot and cold nuclear matter are  of finite range
$r_{int.}=\mu^{-1}$ and we shall assume that $\lambda_q \mu \gg 1$, where
$\lambda$ is the quark mean free path.

\begin{figure}[b!]
\includegraphics[width=3.2in,height=1.in,angle=0]{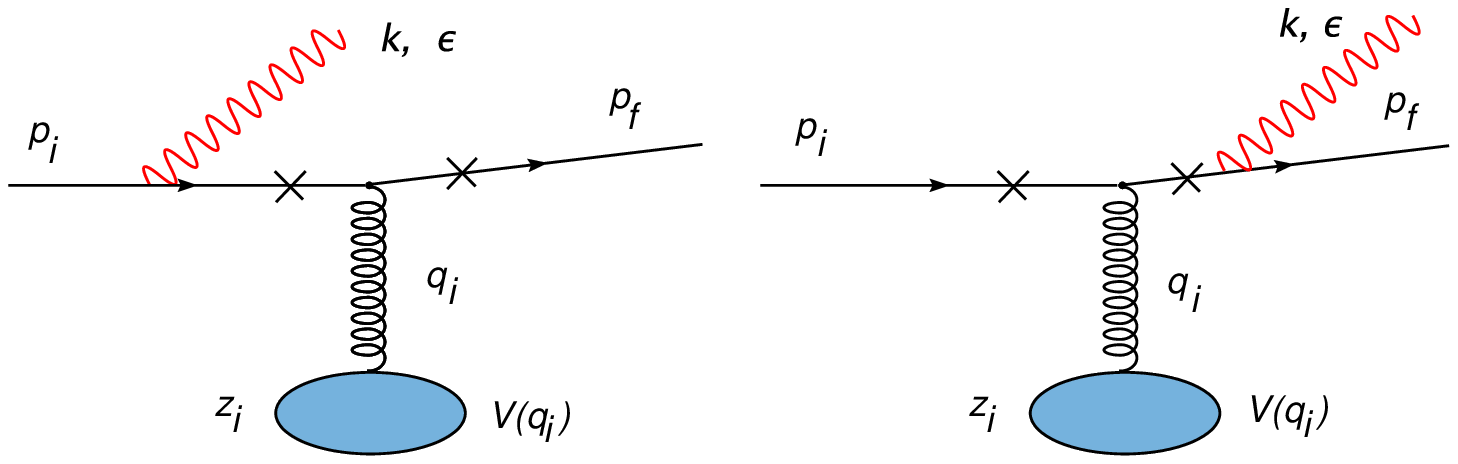} \\
\vspace*{.2cm}
\includegraphics[width=3.2in,height=1.in,angle=0]{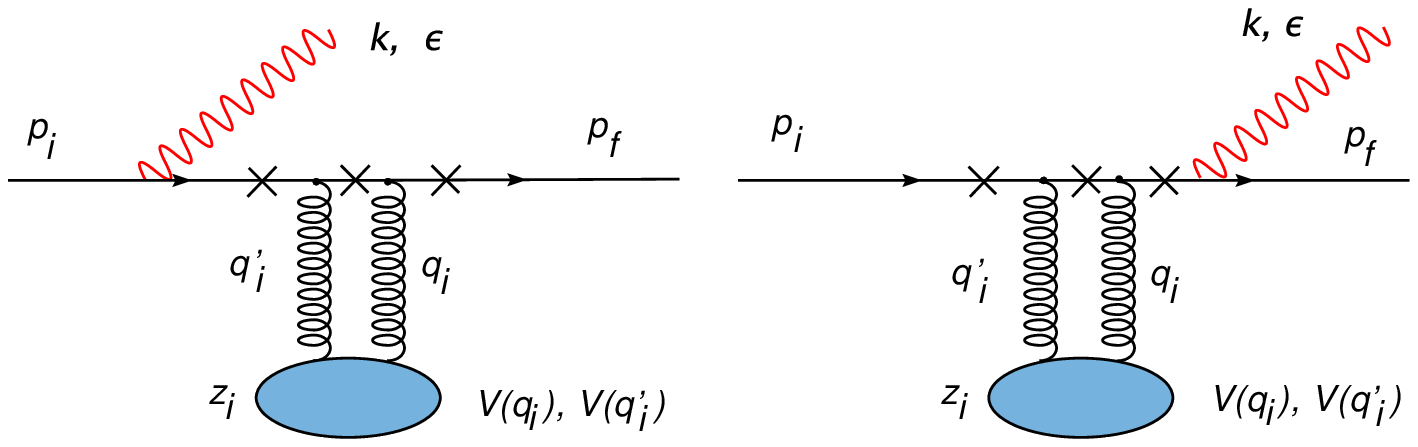}
\caption{ Top panel: single-Born diagrams for medium-induced $\gamma$
emission. Bottom panel: the corresponding double-Born diagrams in
the $z_{i'} \rightarrow z_i$ limit. The third diagram, known to
vanish in this limit~\cite{Gyulassy:2000er}, is not shown. We denote by
`` $  \times $''  the propagators that enter the calculation. }
\label{diagrams}
\end{figure}

The differential photon bremsstrahlung spectrum arises from single-Born
scattering diagrams shown in the top panel of Fig.~\ref{diagrams}.
Using a high energy approximation for the quark to simplify the
interaction and emission vertexes we obtain:
\beqar
i {\cal M}^D_{RHS}(k)  &=&  \int \frac{d^4q}{(2\pi)^4} (-ie)
 \frac{i \epsilon^\mu (2 p_f +k )_\mu}{ (p_f+k)^2 + i \epsilon }
 \nonumber \\
&& \times\;  (-ig_sT^c(p)) \frac{iV^{\nu,c}(q)
(2p_f+2k-q)_\nu }{(p_f+k-q)^2 + i\epsilon}
\nonumber \\
&\approx& \left[ -i \int \frac{d^2 {\bf q}_\perp}{(2\pi)^2}
 v({\bf q}_\perp ) e^{-i {\bf q}_\perp \cdot {\bf z}_\perp }
T^c(p) T^c(t) \right]
\nonumber \\
&& \times \, e\left( \frac{\epsilon \cdot p_f}{k \cdot p_{f}} \right)
e^{iz^+k^-} \; .
\eeqar{ReRHS}
For the second $\gamma$ emission diagram similar
considerations lead to:
\beqar
i {\cal M}^D_{LHS}(k)&\approx&
\left[ -i \int \frac{d^2 {\bf q}_\perp}{(2\pi)^2}
 v({\bf q}_\perp ) e^{-i {\bf q}_\perp \cdot {\bf z}_\perp }
T^c(p) T^c(t) \right]
\nonumber \\
&& \times \; e\left( -\frac{\epsilon \cdot p_i}{k \cdot p_i} \right)
e^{iz^+k^-} \; ,
\eeqar{ReLHS}
and the radiative matrix element at position $z_i$ reads:
\begin{equation}
{\cal M}_{rad}(k,\{i\}) = e\left( \frac{\epsilon \cdot p_f}{k \cdot p_{f}} -
\frac{\epsilon \cdot p_{i}}{k \cdot p_{i}}  \right) e^{i  z_i^+ k^-} \; .
\label{hard}
\end{equation}
In Eq.~(\ref{hard}) the collisional amplitude is not shown.
Let the initial- and final-state momenta of a fast on-shell quark
be $p_i = [ E^+,
{Q}_{\perp \; i-1}^2/(2E^+), {\bf Q}_{\perp \; i-1} ]$, $p_f = [
E^+,  {Q}_{\perp \; i}^2/(2E^+), {\bf Q}_{\perp \; i} ]$, such that
${\bf Q}_{\perp \; i} - {\bf Q}_{\perp \; i-1} = {\bf q}_{\perp \; i}$.
The double
differential medium-induced photon distribution is then
given by:
\beqar
&& \!\!\!  k^+ \frac{dN^\gamma (k;\{i\})}{dk^+ d^2{\bf k}_\perp} =
\frac{1}{2(2\pi)^3}|{\cal M}_{rad}(k,\{i\}) |^2 \nonumber \\
&&\!\!\! =
\frac{\alpha_{em}}{\pi^2}
 \frac{\left( \frac{k^+}{E^+}\right)^2\, {\bf q}_{\perp \,i}^2}{
\left({\bf k}_\perp - \frac{k^+}{E^+} {\bf Q}_{\perp \; i-1} \right)^2
\left({\bf k}_\perp - \frac{k^+}{E^+} {\bf Q}_{\perp \; i} \right)^2 } \; ,
\quad
\eeqar{medgam}
and is dominated by emission coincident with the directions of the
incoming and the outgoing quarks. Changing variables
${\bf \kappa} = {\bf k}_\perp - {\bf k}_\perp^{(\rm pole)}$,
 ${\bf k}_\perp^{(\rm pole)} =  {\bf Q}_{\perp \; i} k^+/ E^+,
{\bf Q}_{\perp \; i-1} k^+/ E^+ $, respectively, we obtain
the QED double logarithmic result: \beq
N^\gamma(\{ i \}) \approx 2 \frac{\alpha_{em}}{\pi} \ln
\frac{k^+_{\max}}{k^+_{\min} }
 \ln  \frac{q^2_{\max}}{m^2}    \; ,
\eeq{Nphot} where  $m^2$ regulates the collinear divergence.

In the case of coherent gluon emission in finite media with few
subsequent scatterings, the interference between the hard (vacuum) and soft
(medium-induced) bremsstrahlung largely determines the
LPM cancellation pattern~\cite{Gyulassy:2003mc,Gyulassy:2000er}.
Double-Born diagrams, with two momentum exchanges at the same
position $z_i = z_i'$, can contribute at any fixed order in opacity.
The relevant double-Born diagrams for photon emission are
shown in the bottom panel of Fig.~\ref{diagrams}.
We find:
\beqar i {\cal M}^V_{RHS}(k) \!
&=& \! \! \int \frac{d^4q}{(2\pi)^4}  \frac{d^4q'}{(2\pi)^4}
    (-ie) \frac{i \epsilon^\mu (2 p_f +k )_\mu}{ (p_f+k)^2 + i \epsilon }
 \nonumber \\
 &\times & (-ig_sT^c(p)) \frac{iV^{\nu,c}(q) (2p_f+2k-q)_\nu }{(p_f+k-q)^2 +
i\epsilon}
 \nonumber \\
&\times &  (-ig_sT^d(p))  \frac{iV^{\delta,d}(q')
(2p_f+2k-2q-q')_\delta } {(p_f+k-q-q')^2 + i\epsilon}
 \nonumber \\
& \approx & \frac{1}{A_\perp} \left[ - \frac{1}{2} \int \frac{d^2
{\bf q}_\perp}{(2\pi)^2} | v({\bf q}_\perp )|^2  \frac{C_t C_p}{d_A}
\right]
\nonumber \\
&& \times \, e\left( \frac{\epsilon \cdot p_f}{k \cdot p_{f}} \right)
e^{iz^+k^-} \; .
\eeqar{VRHS}
To obtain the result in  Eq.~(\ref{VRHS}) we averaged over the initial
and summed over the final parton colors. We also carried out the
average over the position of the scattering center in the transverse plane:
$A_\perp^{-1} \int d^2 {\bf z}_\perp \exp[-i {\bf z}_\perp \cdot
({\bf q}_\perp +{\bf q}_\perp^\prime ) ]   =  A_\perp^{-1} (2\pi)^2
\delta^2 ({\bf q}_\perp +{\bf q}_\perp^\prime )$. For
single-Born interactions, see Eqs. (\ref{ReRHS}) and (\ref{ReLHS}), such
averages are possible only after squaring the amplitudes. The
result differs from the one   for virtual interactions in yielding
$\delta^2 ({\bf q}_\perp -{\bf q}_\perp^\prime )$,
and in the absence of the factor $-1/2$, which accompanies the
collision term.

Similarly, for the second virtual diagram in Fig. \ref{diagrams} we obtain:
\beqar
i {\cal M}^V_{LHS}(k)
& \approx & \frac{1}{A_\perp} \left[ - \frac{1}{2}
  \int \frac{d^2 {\bf q}_\perp}{(2\pi)^2}
| v({\bf q}_\perp) |^2  \frac{C_t C_p}{d_A} \right]
\nonumber \\
&& \times \, e \left( - \frac{\epsilon \cdot p_i}{k \cdot p_{i}} \right)
  e^{iz^+k^-}  \; .
\eeqar{VLHS}
Adding Eqs. (\ref{VRHS})  and (\ref{VLHS}) we see that the same general
radiation matrix element, Eq. (\ref{hard}),  can be factorized for double-Born
interactions. In Eqs.~(\ref{VRHS}) and (\ref{VLHS})
$d \sigma^{\rm el}/d^2{\bf q}_\perp =
 ({C_t C_p}/{d_A})  | v({\bf q}_\perp |^2/ (2\pi)^2$ is the differential
scattering cross section, calculated  in the Born approximation.
However, ${\bf q}_\perp +{\bf q}_\perp^\prime = 0$, and $p_i=p_f$ implies
that the double-Born interaction does not contribute a new
photon bremsstrahlung amplitude.

\section{Differential photon bremsstrahlung spectrum
to all orders in opacity}
\label{allord}

To define the iterative procedure of computing the photon
bremsstrahlung contribution from multiple scattering, we first
consider the action of the direct operator $\htD_n$ at position $z_n$
on a radiative amplitude with $n-1$ correlated scatterings. In what
follows we have dropped the collisional amplitudes since they
were shown to yield an elastic scattering cross section per
order in opacity for both diffusion~\cite{Qiu:2003pm,Adil:2006ra} and
radiative~\cite{Gyulassy:2000er} processes.
The result of such  action can be represented as:
\beqar
&&   \htD_n \vAim({ k}) \; \equiv \;
( \hat{1} + \htB_n) \vAim({ k}) \nonumber \\[1ex]
&& =  \vAim({ k})    + \left(-\half \,\right )^{N_v(\vAim)} {\cal
M}_{rad}(k,\{n\}) \;. \qquad
\eeqar{didamit}
Here, the factor $ \left(-\half \,\right )^{N_v}$ arises because
every virtual contact
interaction in the amplitude gives a factor $-\half$, and
$N_v(\vAim)$ is their number.  The
first term in Eq.~(\ref{didamit}) corresponds to a momentum exchange
with the energetic jet. In the high energy limit  we do not
keep track of the transverse modification of the parent parton
except for the contribution to the soft photon bremsstrahlung. This
approximation does not affect the calculation of the intensity
spectrum $dI^\gamma/dk^+$ but the angular distribution must be
convoluted with the medium-induced jet acoplanarity, which we here
neglect. The second term in Eq.~(\ref{didamit}) is the new radiative
contribution, Eq.~(\ref{hard}), at position $z_n$, and the prefactor
accounts for the number of preceding virtual interactions in the
amplitude.
Next, we consider the double-Born interaction of the quark
at position $z_n$. From section~\ref{zero} we know that there is
no new $\gamma$ bremsstrahlung contribution since $p_f=p_i$.
However,  a factor  $-1/2$ arises for the forward
elastic scattering. The modification of the amplitude
$\vAim(k)$ is found to be:
\beq
 \htV_n \vAim({ k})
=  - \frac{1}{2}  \vAim({k})
\;. \;\;
\eeq{vidamit}
In both Eq.~(\ref{didamit}) and Eq.~(\ref{vidamit}) we have omitted
the color factors since for the simple case of individual parton
propagation these are trivially absorbed in the elastic scattering cross
section or inverse mean free path per order in opacity~\cite{Qiu:2003pm}.

Let $i_n = 0, 1, 2$ indicate for each $n$: no interaction,
direct interaction and
virtual interaction with the medium, respectively~\cite{Gyulassy:2000er}.
Expansion in powers of $\sigma^{\rm el}$ or, equivalently,
$1/\lambda_q$ requires that in the conjugate amplitude ${\vAbi}({ k})$
we have $\bar{i_n} = 2 -i_n $. Consequently, the contribution
to the radiation pattern at $n$-th order in opacity is:
\beqar
&& \!\!\!\!\!\! dN^\gamma(k,n) \propto  \sum_{i_1 \cdots i_n = 0 }^2
{\vAbi}({ k}){\vAi}({ k}) \nonumber  \\
&& \vspace*{-2cm}  \!\!\!\!\!\!  = \sum_{i_1 \cdots  i_{n-1} = 0 }^2
{\vAbim}({ k})( \htD^\dagger \htD + \htV^\dagger + \htV ){\vAim}({ k})
\nonumber  \\
&&  \!\!\!\!\!\!   = \sum_{i_1 \cdots  i_{n-1} = 0 }^2
{\vAbim}({ k})( \hat{B}_n^\dagger \hat{B}_n
+ \hat{B}_n^\dagger + \hat{B}_n ){\vAim}({ k})  \;. \nonumber \\
\eeqar{radpat}
For medium induced photon emission, the reaction operator
$\htR_n =  \htD_n^\dagger  \htD_{n} +  \htV_{n}^\dagger +
\htV_{n} =  \hat{B}_n^\dagger \hat{B}_n + \hat{B}_n^\dagger +
\hat{B}_n$
has a particularly simple form. The first term in
Eq.~(\ref{radpat}) vanishes beyond first order ($n=1$)  in opacity
\cite{Gyulassy:2000er}:
\beqar &&\sum_{i_1 \cdots  i_{n-1} = 0 }^2
{\vAbim}({ k}) \hat{B}_n^\dagger \hat{B}_n {\vAbim}({ k})
 \nonumber \\
&& = |{\cal M}_{rad}(k,\{n\})|^2  \sum_{i_1 \cdots  i_{n-1} = 0 }^2
\left(-\half \,\right )^{\bar{N}_v} \left(-\half \,\right
)^{N_v}= 0\;, \qquad
\eeqar{radpat1}
where we have used $ \left(-\half -\half + 1 \right)^{n-1}= 0$ for
$ n \geq 2 $~\cite{Gyulassy:2000er}. The 2$^{\rm nd}$ and
3$^{\rm rd}$ terms in Eq.~(\ref{radpat}) yield: \beqar && 2 Re \,
{\cal M}_{rad}^*(k,\{n\}) \sum_{i_1 \cdots i_{n-1} = 0 }^2
\left(-\half \,\right )^{\bar{N}_v} {\vAbim}({ k})
 \nonumber \\
&& = \; 2 Re \, {\cal M}_{rad}^*(k,\{n\}) {\cal M}_{rad}(k,\{n-1\})
\nonumber \\  && \qquad \times  \sum_{i_1 \cdots  i_{n-2} = 0 }^2
 \left(-\half \,\right )^{\bar{N}_v}
\left(-\half \,\right )^{N_v} = 0\;, \qquad
\eeqar{radpat2}
if $n\geq 3$. Therefore, unlike the case of gluon bremsstrahlung,
photon bremsstrahlung contributions vanish beyond second order in
opacity.

Taking into account the interactions of the parent quark along its
trajectory through the QGP and the average over the transverse momentum
transfers, the main theoretical result derived in this letter reads:
\beqar
&&k^+ \frac{dN^\gamma (k)}{dk^+ d^2{\bf k}_\perp}
=  \frac{\alpha_{em}}{\pi^2} \bigg\{ \int\frac{d \Delta z_1}{\lambda_q(z_1)}
\int d^2 {\bf q}_{\perp \, 1}
\frac{1}{\sigma^{\rm el}}\frac{d^2 \sigma^{\rm el}}{ d^2 {\bf q}_{\perp \,1 } }
\nonumber \\
&& \times \left[ |{\cal M}_{rad}(\{1\})|^2 +2 {\cal M}^*_{rad}(\{1\})
{\cal M}_{rad}(\{0\}) \cos( k^- \Delta z_1^+) \right] \nonumber \\
&& + \prod_{i=1}^2 \left[  \int\frac{d \Delta z_i}{\lambda_q(z_i)}
\int d^2 {\bf q}_{\perp \, i}  \frac{1}{\sigma^{\rm el}}
\frac{d^2 \sigma^{\rm el}}
{ d^2 {\bf q}_{\perp \,i } }  \right]   \nonumber \\
&& \times \, 2 {\cal M}^*_{rad}(\{2\})
{\cal M}^*_{rad}(\{1\}) \cos( k^- \Delta z_2^+) \bigg\}    \;\;.
\eeqar{prres}
In Eq.~(\ref{prres}) $\Delta z_i^+ = z_i^+ - z_{i-1}^+$, the
$\Delta z_i$ integrals are nested, and
$ \tau_f^{-1} =  {\bf k}^2/(2 \omega) \approx  \surd{2}\, k^-$ is the inverse
photon formation time. When $ \tau_f^{-1} \lambda_q \gg 1$ the photons decohere
early from the parent quark and our result reduces to incoherent emission
from individual scattering centers. Our result differs from
previous findings~\cite{Zakharov:2004bi,AMY:2001} in several important
ways: first, it treats the case of finite  $L/\lambda_q \sim {\rm few}$,
relevant  to heavy ion physics. Second, we find that the most
significant contribution  to the LPM effect for photons
comes from the interference of the medium-induced
photon radiation with the hard emission from the large $Q^2$
scattering of  the parent quark. Finally,
we find that there can be non-linear corrections $\sim L^2$ to the
dominant linear in $L$ behavior of the photon spectrum.
Note that calculations of final-state $\gamma$ emission
have also been carried out in deep inelastic scattering on
nuclei~\cite{Zhang:2003,Majumder:2007ne}
using the high twist  approach~\cite{Wang:2001ifa}.

\begin{figure}[t!]
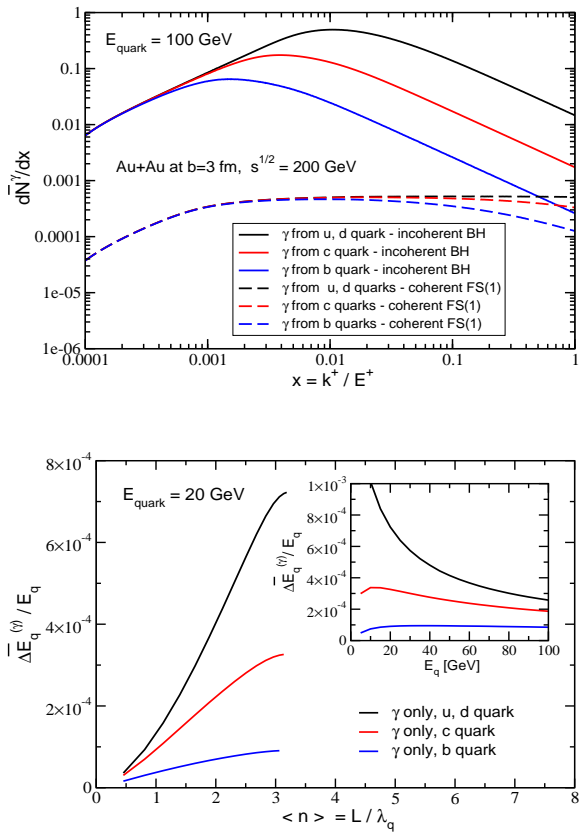

\includegraphics[width=3.in,height=2.in,angle=0]{sptr.eps} \\
\vspace*{.8cm}
\includegraphics[width=3.in,height=2.in,angle=0]{eldep.eps}
\caption{ Top panel: medium-induced photon number spectrum
versus $x=k^+/E^+$ for $E_q = 100$~GeV light, charm and bottom quarks
in central Au+Au collisions at $\sqrt{s} = 200$~GeV. Both the
incoherent Bethe-Heitler (solid) and the coherent final-state
(dashed) bremsstrahlung cases are shown. Bottom panel: partial
fractional quark energy loss $\Delta E_q^{(\gamma)} / E_q$
due to $\gamma$ emission versus the mean number of quark
interactions $\langle n \rangle = L/\lambda_q$. Insert illustrates
the energy dependence of $\Delta E_q^{(\gamma)} / E_q$. }
\label{fig-sptr}
\end{figure}

We are now ready to study numerically the final-state QGP-induced photon
spectrum. Our  results are limited to first order in opacity.
The question of whether destructive interference may even cancel part
of the hard photon bremsstrahlung is deferred  for future studies.
With Eq.~(\ref{hard}) representing the amplitude of both  light and
heavy fermions, the intensity spectrum is easily generalized
to quarks with physical and thermal mass, $m^2 = M_q^2 + C_F g_s^2 T^2 / 4$.
Notable differences from the $m=0$ case include regulation of the poles
in Eq.~(\ref{medgam}) via $ (k^+/E^+)^2 m^2$, appearance of new terms
$\propto m^2$, and finite quark velocity, $\beta < 1 $. We show an
example of a quark propagating outwards from the center of the medium
created in $b=3$~fm Au+Au collisions at RHIC. For a typical integrated gluon
rapidity density,  $dN^g/dy \simeq 1150$,  distributed proportional to
the 2D participant number density  $dN_{\rm part}/d^2{\bf x}_\perp$, we
calculate the temperature $T({\bf x}_\perp , t)$ as a function of time
and position in the transverse plane, assuming approximate Bjorken
expansion of the QGP near midrapidity $y=0$. The necessary Debye screening
scale $\mu(T)=m_D(T)=g_sT$, elastic scattering cross section $\sigma^{qg}(T)$, and
quark mean free path $ \lambda_q(T) = 1 / \sigma^{qg}(T)\rho(T)$
are evaluated as in~\cite{Vitev:2005he} with $g_s=2.5$. The top panel
of Fig.~\ref{fig-sptr} shows the medium-induced photon number spectrum
$\bar{ d {N}^\gamma}/dx = (e/e_q)^{2}  d{N}^\gamma/dx $, $x = k^+ / E^+$,
normalized by the squared fractional quark electric charge
(bar will denote such scaling for any physics quantity).
We considered light, $M_{u,d}=0$~GeV, and heavy, $M_{c}=1.5$~GeV,
$M_{b}=4.5$~GeV, quarks of energy $E_q = 100$~GeV in the Bethe-Heitler
limit and to first order in opacity. In the absence of coherence, the
induced $\gamma$ spectrum is dominated by photon energies $\omega \sim$
a few $\times T$. The contributions of the heavy quark sector to the
medium-induced photon multiplicity and  the energy loss
due to photon emission are strongly suppressed. When the interference
between the vacuum and the medium-induced photon radiation is taken
into account, we find that the spectrum $dN^\gamma/dx$
is suppressed
much more effectively than in the limit of very large number of soft
scatterings for on-shell quarks~\cite{AMY:2001}. The bottom
panel of  Fig.~\ref{fig-sptr} shows
the dependence of the partial fractional energy loss due to $\gamma$
emission versus the mean number of quark interactions in the QGP:
\beq
\langle n \rangle = L / \lambda_q = \int dz / \lambda_q(z) \;.
\eeq{meanopac}
We note that for light and moderately heavy quarks, such as the
charm quark, there is clear non-linear dependence of
$\Delta E_q^{(\gamma)} / E_q$ on  $ \langle n \rangle $. It is
correlated with  deviations from the naive expectation,
$\Delta E_q^{(\gamma)} \propto  E_q$, as illustrated in the insert in
Fig.~\ref{fig-sptr}. Medium-induced photon emission in both the coherent
and incoherent limits is too small, $\Delta E_q^{(\gamma)} / E_q < 1\%$,
to contribute to the quenching of
quark jets. Last but not least, we emphasize that the number of
interactions in the QGP, even for jets emerging from the center of the
heavy ion collision region, is small, $ \langle n \rangle = 3.1$.
As in the case of gluon emission, this is a clear indication
that, even in the most central Au+Au reactions at RHIC, we
are {\em not}  in the limit of large number of scatterings. For
consistency with the calculation of light hadron attenuation,
in our numerical estimates we use an effective geometry of
$L=6$~fm and uniform distribution of partons~\cite{Vitev:2005he}.
In this case  $\langle n \rangle$ is reduced to 2.4,
accounting for jets close to the periphery of the interaction
region.

\vspace*{.5cm}

\section{ Phenomenology of hard photon production in
          p+A and A+A collisions }

\label{study}

\begin{figure}[b!]
\includegraphics[width=3.in,height=2.2in,angle=0]{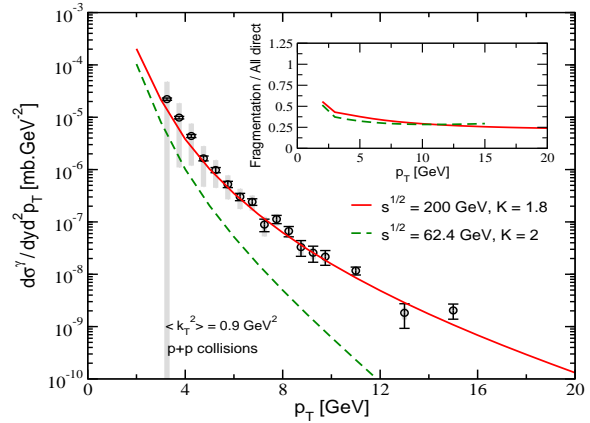} \\
\caption{ Direct photon production cross section in p+p collisions
at $\sqrt{s} = 62.4$~GeV  and 200~GeV. Data at the higher RHIC energy is
from PHENIX~\cite{Adler:2006yt}.  Insert shows the fraction of
fragmentation to all direct photons.}
\label{fig-LOgamma}
\end{figure}

With the results from the previous section at hand, we now turn
to the question of hard, $ p_T > $~a few~GeV, photon production in
heavy-ion collisions. QGP and cold nuclear matter effects are
identified through the nuclear modification ratio:
\begin{eqnarray}
R_{AB}(p_T,b) = \frac{d\sigma_{AB}}{dyd^2{\bf p}_T}
\bigg/ N_{AB}^{\rm coll}(b)\frac{d\sigma_{pp}}{ dyd^2{\bf p}_T } \;,
\label{rab}
\end{eqnarray}
where the number of binary collisions, $N_{AB}^{\rm coll}(b)$ in
Eq.~(\ref{rab}), is computed in an optical Glauber model. The
baseline p+p cross section  is evaluated in factorized perturbative
QCD  to lowest order and leading twist as follows:
\begin{eqnarray}
\label{single}
\frac{ d\sigma^{\gamma}_{pp} }{ dy  d^2{\bf p}_T  }  & = &
K  \sum_{abcd}  \int  dy_d \int  d^2 {\bf k}_a  d^2 {\bf k}_b \;
\frac{ f({ k}_{b}) f({k}_{b}) }{|J({k}_a,{k}_b )|}
\nonumber \\
&& \times \,
 \int \frac{dz}{z^2} \, D_{\gamma/c}(z,\mu_{fr}) \,
\frac{\alpha_s(\mu_r) \alpha_{c} }{2S  }
  |\overline {M}_{ab\rightarrow cd}|^2
\nonumber \\
 &&  \times   \frac{\phi_{a/N}({x}_a,\mu_f)
\phi_{b/N}({x}_b,\mu_f) }{{x}_a{x}_b} \,  \;. \qquad
\label{LO}
\end{eqnarray}
In Eq.~(\ref{LO}) we adhere to the standard notation, see for example
\cite{Owens:1987,Vitev:2005he,Qiu:2004da}, and the Jacobian reads:
\beq
J_{x_a,x_b} ({k}_a,{k}_b ) =
\frac{S}{2}\left(1-\frac{k_a^2k_b^2}{x_a^2S x_b^2S} \right) \; ,
\eeq{jac}
subject to the hard scattering constraint $k_{a,b} < x_{a,b} \sqrt{S}$.
In elementary p+p collisions we use
$\langle { k}_{a,b}^2 \rangle = 0.9$~GeV$^2$ for the normalized
Gaussian parton transverse momentum distributions, $f({ k}_{a,b})$,
and phenomenological K-factors cancel in $R_{AB}(p_T)$. One observes
that Eq.~(\ref{LO}) includes both prompt photons, $c =  \gamma$,
$\alpha_c = \alpha_{ em} = 1/137.036$,
$D_{\gamma/\gamma}(z) = \delta(1-z)$, and fragmentation photons,
$c = q,\bar{q},g$, $\alpha_c = \alpha_{s}(\mu_r)$,
with $D_{\gamma/c}(z,\mu_{fr})$ taken from Ref. \cite{Owens:1987}.
Figure~\ref{fig-LOgamma} shows the cross sections for direct $\gamma$
production in p+p collisions at $\sqrt{s} = 62.4$~GeV and 200~GeV
and identifies the fragmentation fraction of direct photons
that will undergo final-state modification in the QGP.

The first and necessary step, that has thus far been neglected in
direct photon phenomenology, is a systematic study of nuclear
effects in p+A reactions. Not only do the isospin effect, Cronin
effect, shadowing,  and energy loss in nuclei significantly affect
the  observable $R_{AA}(p_T)$,  but these are interesting in their
own right in light of the current and future d+Au measurements at
RHIC~\cite{Peressounko:2006qs}. Theoretical approaches to the Cronin
effect are well documented~\cite{Accardi:2002ik} and are centered
around the idea of initial-state multiple
scattering~\cite{Vitev:2003xu}. In our calculations the Cronin
effect in the nuclear medium is controlled by the broadening of the
transverse momenta of incoming partons  as follows:
\begin{eqnarray}
\langle k_T^2 \rangle &=& \langle k_T^2 \rangle_{pp}
 + \langle k_T^2 \rangle_{med} \; , \nonumber \\
 \langle k_T^2 \rangle_{med} &=&
 \left(\frac{2\mu^2 L}{\lambda}\right)_{q,g}\times 
\max [ 1,\ln(1+\delta \,p_T^2) ] \;, \qquad
 \label{eq:Cronin}
\end{eqnarray}
where the term $\ln(1+\delta \, p_T^2)$ takes into account the power law
tail of the $k_T$ distribution beyond the naive Gaussian random walk
\cite{Vitev:2002pf} (e.g. Moliere multiple scattering). Specifically, 
$\mu^2 = 0.12$~GeV$^2$, $\lambda_g = (C_F/C_A) \lambda_q = 1$~fm 
and $\delta = 0.3$ are fixed to reproduce the experimental p+A 
data when we include initial-state energy loss and are compatible
with our earlier findings for  $\langle k_T^2 \rangle_{med}$  within 
20\% \cite{Vitev:2002pf,Vitev:2003xu}. 
In Eq.~(\ref{eq:Cronin})  $L$ gives the length of
the nuclear medium. We note that in semi-inclusive deep inelastic
scattering (SIDIS) the final-state broadening is expected to be approximately
equal to the initial-state diffusion in $k_T$. A naive estimate of the 
medium-induced contribution to the hadron $ \langle \Delta p_T^2 \rangle_{A} =
 \langle p_T^2 \rangle_{A} - \langle p_T^2 \rangle_{D} $ can be made from 
Eq.~(\ref{eq:Cronin}). The numerical results at leading order for 
 $Q^2 = 5$~GeV$^2$, 
$\langle \Delta p_T^2 \rangle_{Ne} = 0.14$~GeV$^2$ 
 $\langle \Delta p_T^2 \rangle_{Kr} = 0.29$~GeV$^2$ and  $\langle 
\Delta p_T^2 \rangle_{Xe} = 0.36$~GeV$^2$, are  compatible 
within error bars with recent  preliminary HERMES data~\cite{hermes}.

In this work, only the EMC effect was parametrized~\cite{Eskola:1998df} 
and shadowing was calculated from the coherent final-state parton 
interactions~\cite{Qiu:2004da}. The only parameter, the scale 
of power corrections $\xi^2$~\cite{Qiu:2004da}, 
is constrained by the mean squared momentum
transfer per unit length $\mu^2/\lambda$ so that $ (\xi^2
A^{1/3})_{q,g} \approx ( 2 \mu^2 L/ \lambda )_{q,g}$ in minimum bias
reactions, which yields $(\xi^2)_q \approx 0.12$~GeV$^2$. Cold
nuclear matter energy loss \cite{Vitev:2007ve} was consistently 
calculated with the same lengths, momentum transfers and mean free 
paths as in Eq. (\ref{eq:Cronin})  and incorporated in
the pQCD  calculation as follows:
\begin{eqnarray}
\phi_{a,b/N}({x}_{a,b}, Q^2) &\rightarrow & \phi_{a,b/N}
\left( \frac{ x_{a,b} }{1-\epsilon_{a,b}}, Q^2 \right) \nonumber \\
&\approx &
\phi_{a,b/N}\left( {{x}_{a,b}}(1 + \epsilon_{a,b}), Q^2 \right) \; ,\qquad
\label{consist}
\end{eqnarray}
when $\epsilon_{a,b} \ll 1$. Here, $\epsilon_{a,b}$ are the
fractional energy losses for the incoming partons  $a, b$  evaluated
in the rest frame of the corresponding target nucleus. 
It has been shown for the typical hot QGP created at RHIC  that the
effect of multi-gluon fluctuations of the induced bremsstrahlung 
on the quenching of jets could be mimicked by a simple reduction of the 
mean energy loss  $\epsilon_{eff} = \kappa \, \Delta E / E$, 
where  $\kappa = 0.4-0.5 $~\cite{Gyulassy:2001nm}. 
Though fluctuations of the initial-state cold nuclear matter energy loss 
have not yet been studied, in our calculations we make a similar 
approximation. We note that the weaker falloff with energy  of the 
incoming parton flux, when compared to the hard scattered (additional 
$\sim 1/p_T^4$) final-state quark and gluon distributions, implies 
values of $\kappa$ closer to unity. The specific choice that we make,
$\kappa = 0.7$, was constrained through the evaluation of the 
$\pi^0$ production cross section in Cu+Cu collisions at RHIC~\cite{:2008cx} 
which provided a good description of the experimental data.

\begin{figure}[t!]
\includegraphics[width=3.in,height=3.6in,angle=0]{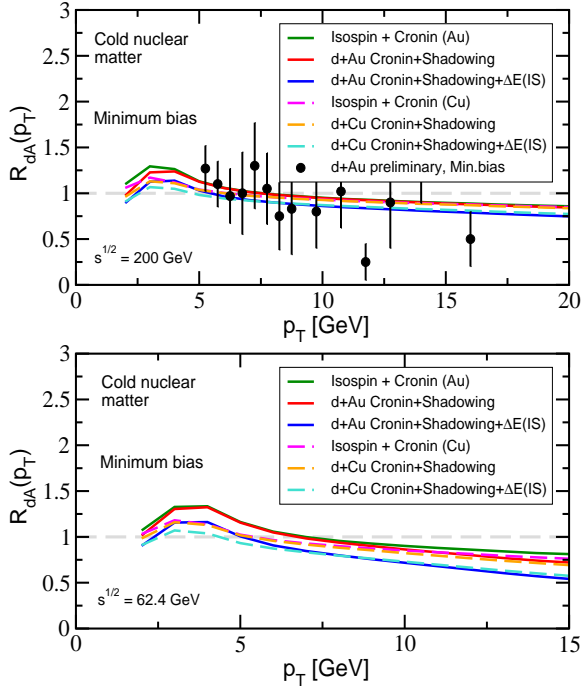} \\
\caption{ Cold nuclear matter effects manifest in direct photon $R_{dAu}(p_T)$
in minimum bias d+Au (solid lines) and
d+Cu (dashed lines) collisions. RHIC energies $\sqrt{s} = 62.4$~GeV
and 200~GeV are shown in the top and bottom panels, respectively.
Preliminary $\sqrt{s} = 200$~GeV minimum bias d+Au data is also
included~\cite{Peressounko:2006qs}.}
\label{fig-dA}
\end{figure}

Our results for minimum-bias d+Au (solid lines) and d+Cu (dashed lines)
collisions are shown in Fig.~\ref{fig-dA}. The calculated $R_{dA}(p_T)$
at $\sqrt{s}=200$~GeV and 62.4~GeV are  presented in the top and bottom
panels, respectively. In the $p_T < 5-7$~GeV region, the nuclear modification
is dominated by Cronin enhancement with a magnitude that is not strongly
affected by shadowing but is sensitive to the initial-state energy loss.
The complementary $p_T > 5-7$~GeV part of phase space
is characterized by  $R_{dA}(p_T)< 1$ with the isospin effect, included
in all calculations, being one of the major contributors. The EMC effect
only  becomes noticeable at the lower C.M. energy and at the highest
transverse momenta. Initial-state, cold nuclear matter energy loss can
contribute as much as the isospin effect at high $p_T$. We emphasize that, at
transverse momenta  $\sim 15$~GeV, nuclear effects on the direct photon
cross section can be as large as 20\% at $\sqrt{s}=200$~GeV and 40\% at
$\sqrt{s}=62.4$~GeV in minimum bias d+A collisions. Preliminary
experimental data~\cite{Peressounko:2006qs} is consistent with the
theoretical expectation. However, given the large error bars, it  cannot
discriminate between different cold nuclear matter effects  or constrain
their magnitudes. Careful experimental investigation is needed to pinpoint
these effects and, since they will be enhanced in A+A collisions,
caution should be exercised in the interpretation of the
$R_{AA}^{\gamma}(p_T)$ findings.

\begin{figure*}[t!]
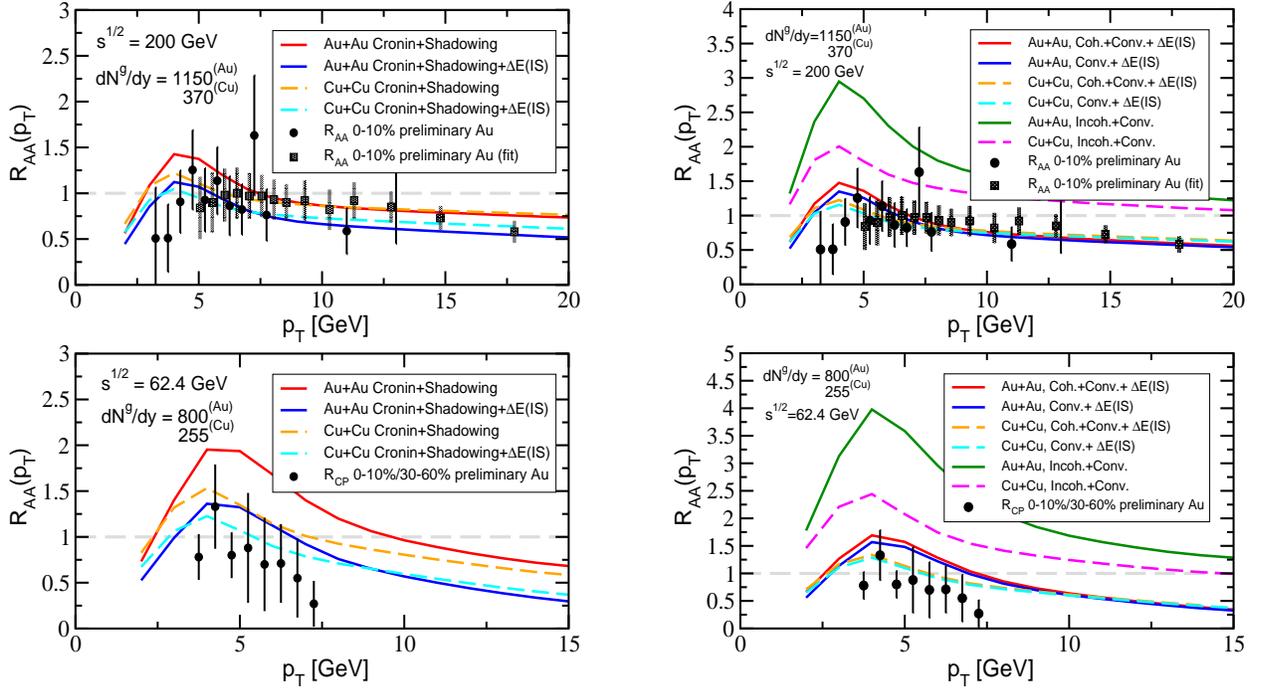

\includegraphics[width=3.in,height=3.6in,angle=0]{gammaAA.eps} \hspace*{1cm}
\includegraphics[width=3.in,height=3.6in,angle=0]{gammaAAfdbk.eps}
\caption{ Calculations of direct photon $R_{AA}(p_T)$ in central
Au+Au (solid lines) and  Cu+Cu (dashed lines) collisions.
RHIC energies $\sqrt{s} = 62.4$~GeV and 200~GeV are
shown in the top and bottom panels. Of the final-state QGP
effects left panels only include the quenching of fragmentation photons.
Right panels also include the jet conversion and medium-induced $\gamma$
(both coherent and incoherent limits) enhancements.  Preliminary
$\sqrt{s} = 200$~GeV and 62.4 GeV Au+Au data is
shown~\cite{Isobe:2007ku,takao}. At 200 GeV two preliminary $R_{AA}(p_T)$
results from~\cite{Adler:2005ig,Adler:2006yt} (circles) and
\cite{takao} (squares) are extracted. }
\label{fig-AA}
\end{figure*}

In A+A collisions, QGP-induced modification of direct photon
production cross section includes competing effects: the quenching
of fragmentation photons and the tree level (jet conversion) and
bremsstrahlung photon enhancement. The quenching of fragmentation
photons is modelled
in the same way as the quenching of hadrons~\cite{Vitev:2003xu}
and can be combined with the medium-induced $\gamma$ contributions
as follows:
\begin{eqnarray}
D_{\gamma/c} (z) & \Rightarrow & \int_0^{1-z} d\epsilon \; P(\epsilon)  \;
\frac{1}{1-\epsilon} D_{\gamma/c} \left( \frac{z}{1-\epsilon} \right)
\nonumber \\  &&
+  \,  \frac{dN^\gamma_{\rm med.}(c)}{d z}  \, + \,
N^\gamma_{\rm conv.}(c) \delta(1-z) \;. \quad
\label{mod1}
\end{eqnarray}
Here, $P(\epsilon)$ is the probability distribution of the fractional
jet energy loss $\epsilon = \Delta E / E$~\cite{Vitev:2005he,Gyulassy:2001nm},
$dN^\gamma_{\rm med.}(c)/{d z}$ is the QGP-induced bremsstrahlung
that we calculated in section~\ref{zero},
and  $N^\gamma_{\rm conv.}(c)$ is the number of jets
converted to photons.
For such conversions, in Eq.~(\ref{mod1}) the $p_\gamma \approx p_c$
approximation~\cite{Fries:2003} that stems from the limit of small
$t$-channel momentum transfers to energetic jets is implicit.
For a parent quark propagating through the QGP:
\beq
N^\gamma_{\rm conv.}(c) =  \int_{t_0}^L dt \;
\rho(T) \sigma^{qg\rightarrow \gamma q}_{tot} (T) \;,
\eeq{Nconv}
where time and position dependence is taken via  $T({\bf x}_\perp, t)$.
The cross section in Eq.~(\ref{Nconv}), with $s \approx 2 m_D E$ and
$t \in (m_D^2, s/4)$, consistent with the forward scattering approximation,
reads:
\beq
\sigma^{qg\rightarrow \gamma q } =
\frac{\pi \alpha_s \alpha_{em}}{6 m_D E}  
\ln\frac{E}{2m_D} \; .
\eeq{totcsec}

Numerical results in central Au+Au and Cu+Cu collisions at the
intermediate and top RHIC energies are shown in
Fig.~\ref{fig-AA}. The final-state gluon rapidity densities
$dN^g/dy = 1150\; (Au,200\; {\rm GeV})$, $800 \; (Au,62.4\; {\rm GeV})$,
$370 \;  (Cu,200\; {\rm GeV})$, $255 \; (Cu,62.4\; {\rm GeV})$ are
constrained by measured and extrapolated particle multiplicities at
RHIC~\cite{Vitev:2005he}. Of the final-state effects in the QGP, the
left panels only include the quenching of fragmentation photons.
Since a small fraction of the direct $\gamma$ come from fragmentation
processes and quark attenuation is significantly smaller, $\sim C_F/C_A$,
when compared to gluon attenuation, the observable QGP modification is
also very small. In fact, $R_{AA}(p_T)$ is dominated by cold nuclear 
matter effects, such as  the ones shown in
Fig.~\ref{fig-dA}, amplified by the presence of two large nuclei.
$R_{AA}(p_T)$ from published~\cite{Adler:2005ig,Adler:2006yt} and
preliminary~\cite{Isobe:2007ku,takao} data are shown for comparison.
Given the error bars, only large Cronin enhancement at $\sqrt{s}= 62$~GeV
is excluded, suggestive of the role of initial-state inelastic jet
scattering  in controlling the magnitude of the Cronin effect.

The right panels in Fig.~\ref{fig-AA} include the medium-induced
photons and the jet conversion contribution. For completeness, we
have also shown results for the incoherent Bethe-Heitler
radiation spectrum and absence of initial state energy loss. This
scenario gives direct $\gamma$ enhancement as large as factors of
3 and 4 over the p+p baseline at $\sqrt{s}=200$~GeV and 62 GeV,
respectively. Not only is this case not supported theoretically by
the results derived in this paper and in Ref.~\cite{Vitev:2007ve},
but when compared to data, even with the present large experimental
error bars, it is clearly excluded. In calculating the coherent
final state photon emission rate, Eq.~(\ref{prres}), and the jet conversion
rate,  Eq.~(\ref{Nconv}),  we also account for the time-dependent
quenching of the quark as it propagates through the QGP:
\beq
R_{AA}(p_T,t) = (1-f(t)) + R_{AA}(p_T) f(t) \;.
\eeq{rateqt}
Here, $R_{AA}(p_T)$ is the full final-state quark quenching,
evaluated as in~\cite{Vitev:2005he} for different system sizes
and center of mass energies, and $f(t)$ interpolates between 0
and 1 to give the time dependence of radiative energy
loss~\cite{Vitev:2007ve}. We find that this effect reduces
$dN^\gamma_{\rm med}/dz$ and $N^\gamma_{\rm conv.}$
by $\sim30\%$.  Our results show that at transverse momenta
$p_T < 5$~GeV the contribution from jet conversion to the
total photon cross section is limited to $\sim 25\%$ and the
contribution of medium-induced $\gamma$ is limited
to  $\sim 10\%$. In the high $p_T$ range the total enhancement
contribution is found to be  $ \sim 5 \%$.

\section{Conclusions}
\label{conclude}

In this paper, we provided a theoretical derivation of the
final-state QGP-induced photon bremsstrahlung for the experimentally
relevant case of hard jet production. We demonstrated that while the
physics processes that control photon and gluon emission differ, the
common Landau-Pomeranchuk-Migdal interference between the radiation
from the hard scattering and the radiation from the subsequent soft
quark interactions in the plasma leads to a significant suppression
of the $\gamma$ intensity. The photon spectrum was found to be
attenuated at least by a factor of several for jet energies relevant
for RHIC phenomenology, in contrast with the estimated modest 30\%
attenuation for asymptotic $t = -\infty$ on-shell jets in the limit
of a very large number of soft interactions, where the interference
with the bremsstrahlung that accompanies hard jet production has not
been taken into account. We found that the suppression of $dI^\gamma
/dk^+$ also leads to non-linear dependence of  $\Delta E_q^\gamma (
L/\lambda_q , E_q)$ on the system size and sub-linear dependence on
the parent quark energy that have previously been associated
primarily with gluon emission.

To help identify the significance of both cold and hot nuclear
matter effects on direct photon production we carried out the first
systematic phenomenological study of $R_{AB}^\gamma (p_T)$ in
midrapidity d+Cu, d+Au, Cu+Cu and Au+Au reactions at RHIC energies
of $\sqrt{s} = 62.4$~GeV and 200~GeV. As expected, in all cases we
found that in the absence of QGP formation the nuclear modification
factor at intermediate transverse momenta is dominated by the Cronin
effect and at high transverse momenta by isospin effects and
initial-state parton energy loss. Surprisingly, however, the
contribution of final-state QGP effects to the direct photon cross
section in nucleus-nucleus collisions was small: less than
$ -25\%$ from the quenching of fragmentation photons,
less than $ +25\%$  from jet conversion and  less than $ +10\%$
from medium-induced $\gamma$ bremsstrahlung. While
experimental measurements are not yet precise enough to disentangle
such modest effects, they have already put severe constraints on
theoretical models that suggest a dominant role of jet-plasma
interactions in the $p_T \geq 2$~GeV part of phase space. In
particular, PHENIX data is compatible with strong coherent
suppression of the bremsstrahlung $\gamma$, which is the main
theoretical result of this paper.

In conclusion, we suggest that only a systematic study of direct
photons in heavy ion reactions for various system sizes and center
of mass energies will help uncover their full potential both as a
baseline for jet tomography and as an independent probe of nuclear
effects. In this exploration, precision d+A data is critical, since
our theoretical results support the possibility that, in both
proton-nucleus and nucleus-nucleus collisions, cold nuclear matter
effects play a dominant role in altering the cross section for
direct photon production.

\vspace*{.1cm}

{ \bf Acknowledgments:} We thank T. Goldman and T. Sakaguchi for
useful discussions. This research is supported by the US Department
of Energy, Office of Science, under Contract No. DE-AC52-06NA25396
and in part by the LDRD program at LANL, the NNSF of China and the
MOE of China under Project No. IRT0624.

\end{document}